\documentclass[11pt]{article}
\title{Airy kernel with two sets of parameters in directed percolation and random matrix theory.}
\author{Alexei Borodin \footnote{Mathematics 253-37, Caltech, Pasadena, CA 91125, USA; E-mail:borodin@caltech.edu\: and Institute for Information Transmission
Problems, Moscow, Russia}\:\: and Sandrine P\'ech\'e \footnote{Institut Fourier, 100 Rue des maths, 38402 Saint Martin d'Heres, France; E-mail: Sandrine.Peche@ujf-grenoble.fr Temporary address while this work was done: Department of Mathematics,
University of California at Davis,
One Shields Ave., Davis, CA 95616, USA.}}
\usepackage{epsfig}
\usepackage{amsfonts}
\usepackage{amssymb}
\usepackage{amsthm}
\usepackage{amstext}
\usepackage{amscd}
\usepackage{amsmath}
\usepackage{delarray}
\newtheorem{theo}{Theorem}
\newtheorem{prop}{Proposition}
\newtheorem{lemme}{Lemma}
\newtheorem{conjecture}{Conjecture}
\newtheorem{definition}{Definition}
\newtheorem{fact}{Fact}[section]

\theoremstyle{remark}

\newtheorem{remark}{Remark}

\newtheorem{Notationnal remark}{Remark}[section]

\newcommand{\brem}{\begin{remark}}
\newcommand{\erem}{\end{remark}}
\newcommand{\bconj}{\begin{conjecture}}
\newcommand{\econj}{\end{conjecture}}
\newcommand{\bdefi}{\begin{definition}}
\newcommand{\edefi}{\end{definition}}
\newcommand{\bt}{\begin{theo}}
\newcommand{\bfa}{\begin{fact}}
\newcommand{\efa}{\end{fact}}

\newcommand{\et}{\end{theo}}
\newcommand{\bp}{\begin{prop}}
\newcommand{\ep}{\end{prop}}
\newcommand{\bl}{\begin{lemme}}
\newcommand{\el}{\end{lemme}}
\newcommand{\be}{\begin{equation}}
\newcommand{\ee}{\end{equation}}
\newcolumntype{L}{>{$}l<{$}}
\newcommand{\mbE}{\mathbb{E}}
\newcommand{\mbR}{\mathbb{R}}
\newcommand{\mbC}{\mathbb{C}}
\newcommand{\mCC}{\mathcal{C}}

\newcommand{\mbP}{\mathbb{P}}

\begin{document}
\maketitle

\begin{abstract}
We introduce a generalization of the extended Airy kernel with two
sets of real parameters. We show that this kernel arises in the edge
scaling limit of correlation kernels of determinantal processes
related to a directed percolation model and to an ensemble of random
matrices.
\end{abstract}
\section{Introduction and results}

The Airy kernel is one of the most fundamental objects of Random
Matrix Theory. The determinantal random point process governed by
the Airy kernel describes the behavior of the largest eigenvalues of
large Gaussian Hermitian random matrices (a.k.a. GUE -- Gaussian
Unitary Ensemble), see \cite{BowickBrezin}, \cite{Forrester},
\cite{TW-Airy}, last passage time in directed percolation models in
a quadrant \cite{JohShapefluctuations}, asymptotics of the longest
increasing subsequences of random permutations \cite{BaikDeiftJoh},
and it also appears in many other problems whose list is too long to
be included here.

The Airy kernel has a time-dependent version usually referred to as
the {\it extended Airy kernel\/}. Originally obtained in
\cite{PraehoferSpohn} via asymptotics of a polynuclear growth model
in 1+1 dimensions, the extended Airy kernel arises in virtually
every problem where the usual Airy kernel comes up, provided that
the probability measure in question is equipped with a natural
Markov dynamics that preserves the measure. In particular, it
describes the edge scaling limit of Dyson's Brownian motion on GUE, 
the change in the quadrant last passage time when the observation
point moves \cite{JohPNG}, and edge behavior of large random
partitions under the Plancherel dynamics related to the longest
increasing subsequences of random permutations \cite{BO-Plancherel}.
The Extended Airy kernel has also appeared in a
much earlier note
\cite{Macedo}, and we are very grateful to a referee for pointing this out.
%The extended Airy kernel also arises in the edge scaling limit of the Brownian motion model
%corresponding to Wishart ensembles as proposed in \cite{Macedo}, where some external perturbation is also considered. 

In \cite{BaikGBAPeche} it was demonstrated that the Airy kernel is
not stable in the sense that it can be naturally viewed as a point
in a family of Airy-like kernels indexed by a finite set of real
numbers. In the context of GUE, those are the (scaled) eigenvalues
of a deterministic perturbation of finite rank, and in the
percolation context the parameters correspond to a few defective
rows or columns in the quadrant. For Wishart ensembles of random
matrices, the same family of kernels was later obtained in
\cite{DesrosiersForrester}. Time-dependent extensions of kernels
from this family appeared in the recent work \cite{ImamuraSasamoto}
on asymptotics of the totally asymmetric simple exclusion process
(TASEP).

The main goal of this note is to introduce a new Airy-like
time-dependent correlation kernel with two sets of real parameters.
We obtain it as a limit of a directed percolation in a quadrant
which has both defective rows and columns. We also show that the
``static'' version of the kernel arises in the edge scaling limit of
a certain Wishart-like ensemble of random matrices. We were unable
to obtain the extended version via random matrices but we do believe
that it should be possible. Our kernel generalizes all the kernels
mentioned above.\\

Let us describe our results in more detail.

Consider a directed percolation model with exponential waiting times
defined as follows. Let $\pi_1, \ldots, \pi_p, \hat \pi_1, \ldots ,
\hat \pi_p$ be fixed real numbers such that $\pi_i+\hat \pi_j >0$
for any $1\leq i,j \leq p$. Let $W=(W_{ij})_{i,j=1, \ldots, p}$ be a
$p\times p$ array of independent exponential random variables with
$\mbE (W_{ij})=(\pi_i+\hat \pi_j)^{-1}$. For any $1\leq N \leq p$,
we consider the so-called last passage time in this percolation
model: \be \label{def: YNP} Y(N,p):=\max_{P \in \Pi} \sum_{(ij) \in
P}W_{ij}, \ee where $\Pi$ is the set of up-right paths from $(1,1)$
to $(N,p)$. The random variable $Y(N,p)$ has a natural
interpretation in terms of queuing theory, due to the result of
\cite{GlynnWitt}. This is the exit time of the $p$th customer in a
series of $N$ files
 where the service times $W_{ij}$ depend on both the file and the customer.
This random variable also has an interpretation in terms of TASEP,
which is a model of interacting particles on $\mathbb{Z}$. One
starts with the initial configuration
$\eta_0(i)=1_{\mathbb{Z}_-}(i)$, meaning that only the negative
sites are occupied. Then, if the site $i+1$ is unoccupied, the
particle at site $i$ jumps to site $i+1$ after a random waiting
time. The waiting times are independent exponential random variables
whose parameters depend on the particle and the number of jumps
already performed by this particle. One can think of $W_{ij}$ as
of the waiting time of $i$th particle from the right performing the jump
number $j$.\\

We first prove the following result.

Let $X_N$ be a $p\times N$ random matrix with independent complex
Gaussian entries \be \label{def: XN} X_{ij}\sim \mathcal{N}\left(0,
\frac{1}{\pi_i +\hat \pi_j}\right).\ee \bt \label{theo: YNPRMT} Let
$\lambda_1$ be the largest eigenvalue of $X_NX_N^*$. Then, for any
$x$, $$ \mbP (Y(N,p)\leq x)= \mbP (\lambda_1\leq x).$$ \et

%\brem Theorem \ref{theo: YNPRMT} establishes a connection between
%the last passage time in directed percolation model with exponential
%waiting times and a random matrix model, which was only known for
%special cases of the parameters $\pi_i$'s and $\hat \pi_j$'s (see
%\cite{BaikGBAPeche} for instance). \erem

A natural question is then to investigate the above connection as a
process. Consider a sequence of growing random matrices $(X_k)_{k=1,
\ldots, p}$ where $X_{k+1}$ is obtained from $X_k$ by adding one
column with random Gaussian entries (with the appropriate variance).
We can then consider the joint distribution of the largest
eigenvalues of the random matrices $X_kX_k^*$, $1\leq k \leq p.$
Simultaneously, one can consider the joint distribution of the
random variables $Y(k,p)$, $1\leq k \leq p$. Are these joint
distributions the same? We cannot establish that the equality
actually holds, due to the fact that the computation of the joint
eigenvalue distribution of the random matrices $X_kX_k^*$ is not an
easy task.\footnote{As was pointed out to us by Peter Forrester, the
equality can be established in the degenerate case when all $\pi_i$
tend to the same constant using the techniques of
\cite{ForresterRains}, see Appendix to \cite{ForresterNagao}.}
Nevertheless, we can study a determinantal point process which
occurs naturally in both models and obtain a new limiting
correlation kernel, which generalizes the extended Airy kernel.\\
Let $J_1, J_2$ be given integers, and $X=\{x_1, x_2, \ldots,
x_{J_1}\}$, $Y=\{y_1, y_2, \ldots, y_{J_2}\}$ be given sets of real
numbers satisfying $x_i>y_j$ for any $1\leq i\leq J_1$ and any
$1\leq j \leq J_2.$ Let $\gamma$ and $\Gamma$ be the contours
defined on Figure \ref{fig: contoursdefi} below.
\begin{figure}[htbp]
 \begin{center}
 \begin{tabular}{c}
 \epsfig{figure=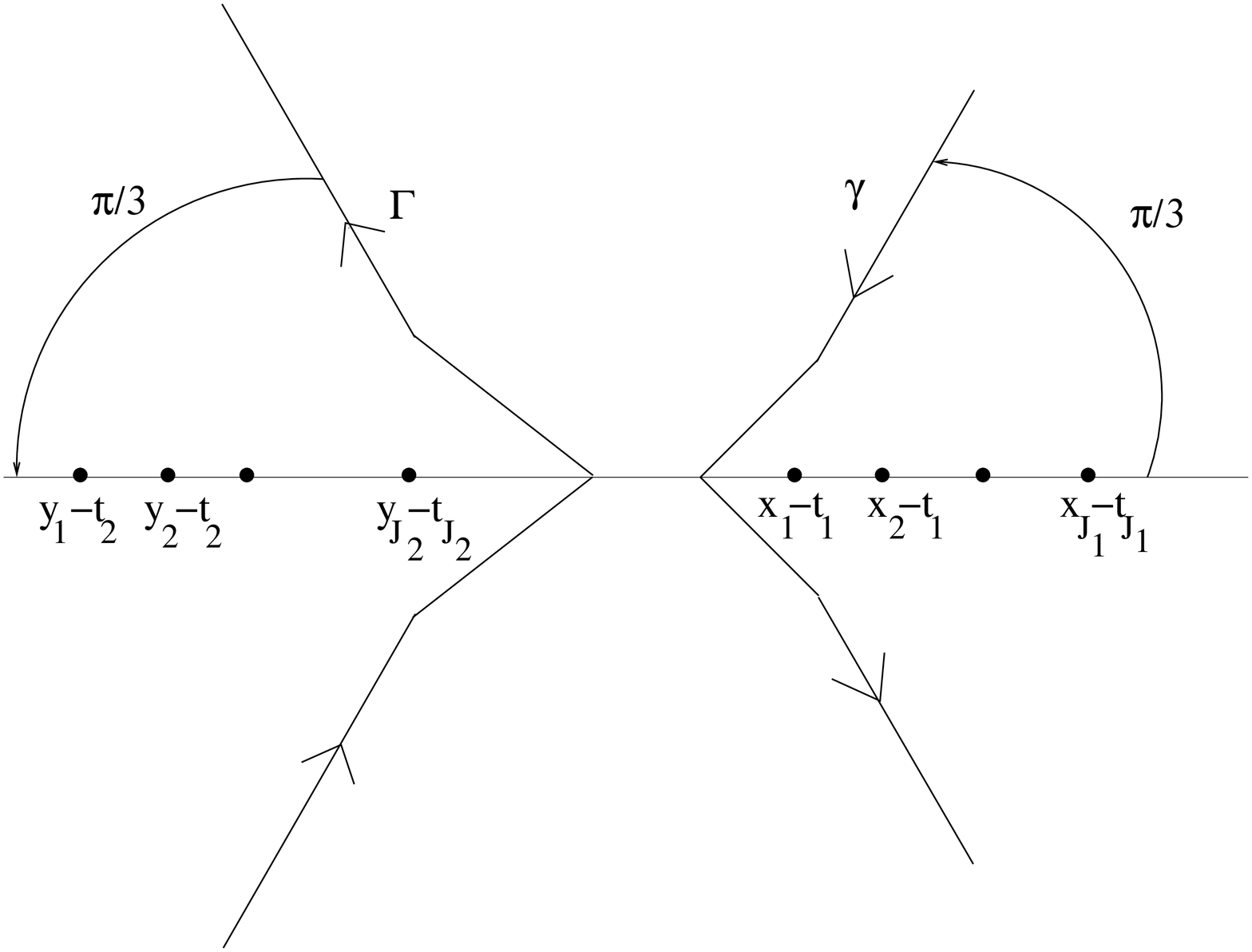,height=6cm,width=7cm, angle=0}
 \end{tabular}
 \caption{The contours of the parameterized extended Airy kernel. \label{fig: contoursdefi}}
\vspace*{-0.3cm}
 \end{center}
\end{figure}

Denote by $K_{Ai}(t_1,x;t_2,y)$ the extended Airy kernel
\begin{eqnarray}\label{def: airykernelextended}
K_{Ai}(t_1,x;t_2,y)=\begin{cases}\displaystyle\int_{0}^{\infty}
e^{-\lambda(t_1-t_2)}Ai(y+\lambda)Ai(x+\lambda)d\lambda\ \text{, if
}\ t_1\ge t_2,\\
\displaystyle -\int_{-\infty}^{0}
e^{-\lambda(t_1-t_2)}Ai(y+\lambda)Ai(x+\lambda)d\lambda\ \text{, if
}\ t_1 <t_2.
\end{cases}
\end{eqnarray}

\bdefi \label{defi: galizedAiry} The extended Airy kernel with two
sets of parameters is defined by
\begin{multline}
K_{Ai; X,Y}(t_1,x;t_2,y)
=K_{Ai}(t_1,x;t_2,y)+ \frac{1}{(2\pi
i)^2} \int_{\gamma}d\sigma\int_{\Gamma}d\tau \\
\frac{e^{y\tau-\tau^3/3-x\sigma+\sigma^3/3}}{\tau+t_2-\sigma-t_1}
\left(\prod_{i=1}^{J_1 }\frac{t_2+\tau-x_i}{t_1+\sigma-x_i}\prod_{i
=1}^{J_2}\frac{ t_1+\sigma-y_i }{t_2+\tau-y_i}-1\right),
\label{extparaAiry2}
\end{multline}
where the two contours $\gamma$ and $\Gamma$ are chosen as on Figure
\ref{fig: contoursdefi}.

\edefi

\brem For any fixed values of $t_1$ and $t_2$, the kernel of
Definition \ref{defi: galizedAiry} is a finite rank perturbation of
the extended Airy kernel. Indeed, by a straightforward computation, one finds that 
%and setting $\tau_2=\tau +t_2$, $\sigma_1=\sigma+t_1$, one finds that
\begin{multline}
K_{Ai; X,Y}(t_1,x;t_2,y)-K_{Ai}(t_1,x;t_2,y)\\=
\frac{1}{(2i\pi)^2} \sum_{k_1+k_2=1}^{J_1+J_2}\int_{\gamma}d\sigma\int_{\Gamma}d\tau\,
e^{y\tau-\tau^3/3-x\sigma+\sigma^3/3} \\ \sum_{1\leq i_1< \cdots < i_{k_1}\leq J_1}\sum_{1\leq j_1< \cdots < j_{k_2}\leq J_2}\frac{(-1)^{k_2}(\tau+t_2-\sigma-t_1)^{k_1+k_2-1}}{\prod_{l=1}^{k_1}(\sigma+t_1-x_{i_l})\prod_{l'=1}^{k_2}(\tau+t_2-y_{j_{l'}})}.
\end{multline}
\erem

\brem One does not have to stop at considering finitely many
perturbation parameters. By taking limits with number of $x_i$'s and
$y_j$'s going to infinity, one arrives at the following kernel. Let
$\{a_i^{\pm}\}_{i=1}^\infty$ and $\{b_i^{\pm}\}_{i=1}^\infty$ be
four sequences of nonnegative numbers such that
$\sum_{i=1}^\infty(a_i^\pm+b_i^\pm)<\infty$, and let $c^{\pm}$ be
two positive numbers. Set \be
\Phi_{a,b,c}(z)=e^{c^+z+c^-z^{-1}}\prod_{i=1}^\infty
\frac{(1+b_i^+z)(1+b_i^- z^{-1})}{(1-a_i^+z)(1-a_i^- z^{-1})}
\label{tp}\,.\ee Then the kernel
\begin{multline*}
K_{Ai;\, a,b,c}(t_1,x;t_2,y)=K_{Ai}(t_1,x;t_2,y)\\+\frac{1}{(2\pi
i)^2}\int_{\gamma}d\sigma\int_{\Gamma}d\tau\, \
\frac{e^{y\tau-\tau^3/3-x\sigma+\sigma^3/3}}{\tau+t_2-\sigma-t_1}
\left(\frac{\Phi_{a,b,c}(\sigma+t_1)}{\Phi_{a,b,c}(\tau+t_2)}-1\right),
\label{extparaAiry2}
\end{multline*}
where the contours are chosen so that all points $1/a_i^+-t_1$ and
$a_i^--t_1$ are to the right of $\gamma$, and all points
$-1/b_i^+-t_2$ and $-b_i^--t_2$ are to the left of $\Gamma$, is
readily seen to be a limit of kernels of Definition \ref{defi:
galizedAiry}. Interestingly enough, functions (\ref{tp}) also
parameterize stationary extensions of the discrete sine kernel, see
\cite{B-Periodic}. They also appear as generating functions of
totally positive doubly infinite sequences \cite{Ed}, and as
indecomposable characters of the infinite-dimensional unitary group,
see \cite{OO-Jack} and references therein. \erem

The main result of this note is the proof that the kernel $K_{Ai;
X,Y}$ arises as a scaling limit of correlation kernels of the
determinantal point processes related to the directed percolation
model in a quadrant, defined as follows.\\
Let $0<t<1$ be a given real number and assume that there exist two integers $J_1, J_2$, real numbers $x_i,1\leq i\leq J_1$, $y_j, 1\leq j\leq J_2$
independent of $p$ such that:
\begin{eqnarray*}
&\pi_i=\dfrac{\sqrt{t}}{\sqrt t +1 }+\dfrac{x_i}{\alpha
p^{1/3}},\quad i=1, \ldots, J_1;& \pi_i=1, i>J_1;\crcr
& \hat
\pi_j=-\dfrac{\sqrt{t}}{\sqrt t +1 }-\dfrac{y_j}{\alpha
p^{1/3}},\quad  j=1, \ldots, J_2,& \hat \pi_j=0, j>J_2.\label{def: param}
\end{eqnarray*}
We consider the associated directed percolation model (\ref{def: YNP}) and prove the following result.\\
Set $$\alpha_s=t+2\frac{\left (t(1+\sqrt t)\right)^{2/3}}{p^{1/3}}s,\, Y_s= Y\left (p\alpha_s,p\right ) \text{ and } \sigma_s=p\left (1+\sqrt{\alpha_s}\right )^2.$$
Let also $\mathcal{A}_s^{ext}$ be the stochastic process whose finite dimensional distributions are given by 
$$\mbP (A_{t_1}^{ext}\leq \xi_1, \ldots, A_{t_m}^{ext}\leq \xi_m)=\det(I-f
K_{Ai; X,Y}f)_{L^2(\Gamma^m \times \mbR^m)},$$
where $f(t_j,x)=1_{x\geq \xi_j}$ and $\Gamma^m=\{t_1, \ldots, t_m\}$. Here $\Gamma^m\times \mbR^m$ is equipped with the product measure $d\mu \otimes d\lambda$ where $d\mu$ is the counting measure and $d\lambda$ denotes the Lebesgue measure.
\bt \label{theo: CVext}  As $p\to \infty$, 
$$p^{-1/3}t^{1/6}(1+\sqrt t)^{4/3}(Y_s-\sigma_s)\to \mathcal{A}_s^{ext},$$
in the sense of convergence of finite dimensional distributions.\et

\paragraph{}The paper is organized as follows. In Sections \ref{Sec: RSK}
and \ref{Sec: RMT}, we study the directed percolation model and its connection
 to the random matrix model defined in (\ref{def: XN}). This connection is
 established using the Robinson-Schensted-Knuth correspondence. In Section
 \ref{Sec: DRPF}, we study the correlation function of the point process
 induced by the joint distribution of $\{Y(k,p)\}_{1\leq k\leq p}$. Last,
  in Section \ref{Sec: asym}, we consider asymptotics of these correlation
  functions thus obtaining the kernel of Definition
  \ref{defi: galizedAiry}.

\paragraph{Acknowledgments.}{The first named author was partially
supported by the NSF grants DMS-0402047 and DMS-0707163. The second
named author was visiting UC Davis when this paper was written. She
would like to thank especially A. Soshnikov who allowed her to spend
a great year at UC Davis and encouraged this work.

The authors are also very grateful to Peter Forrester for a number
of valuable remarks.}
\section{Last passage time in percolation
models with exponential waiting times.
 \label{Sec: RSK}}

\paragraph{}We start with some reminders.
We denote a partition by $\lambda$ throughout the paper. A partition
is an infinite sequence of non-negative integers $(\lambda_1,
\lambda_2, \ldots, \lambda_N, \ldots)$, where $\lambda_1\geq
\lambda_2\geq \dots \geq \lambda_N \geq \cdots$, with finitely many
nonzero entries. The Schur measure introduced in \cite{Ok1} is a
measure on partitions which assigns to a given partition $\lambda$ a
weight as follows. Let $a_1, \ldots, a_N$ and $b_1, \ldots, b_p$ be
given nonnegative real numbers. In what follows we assume that
$a_ib_j<1$ for any $i,j$.

\bdefi \label{defi: schurmeasure} The Schur measure with parameters
$a=(a_1,\dots,a_N)$ and $b=(b_1,\dots,b_p)$ is a probability measure
$\mathcal M$ on partitions assigning to a partition $\lambda$ the
weight \be \label{shurmeasure}{\mathcal
M}(\lambda)=\frac{1}{Z}s_{\lambda}(a)s_{\lambda}(b).\ee  Here $Z$ is
a constant, and $s_\lambda$ denotes the Schur symmetric functions
parameterized by $\lambda$.

\edefi The normalizing constant $Z$ is computed by the well-known
Cauchy identity for Schur functions, which implies
$Z=\displaystyle{\prod_{1\leq i\leq p, 1\leq j\leq N}}(1-a_ib_j).$

\paragraph{}The Schur measure naturally occurs in some directed percolation models, as we now recall.
Let $W=(w_{ij})_{1\leq i \leq p, 1\leq j \leq N}$ be a $p\times N$
random matrix with independent entries with the geometric
distribution:
$$\mathbb{P}(w_{ij}=n)= (1-a_ib_j)(a_ib_j)^n,
\text{ for any $n=0,1,2,\dots$}\,.$$ One can associate to this
random matrix the so-called last passage time in a directed
percolation model defined by \be \label{def:
lastpassagegeo}L(N,p):=\max_{P \in \Pi} \sum_{(ij) \in P}w_{ij},\ee
where $\Pi$ is the set of up-right paths from $(0,0)$ to $(p,N).$
This last passage time can also be understood as the exit time of
the $N$th customer in a series of $p$ files with independent
geometric waiting times of expectation depending on both the files
and customer.

The distribution of the random variable (\ref{def: lastpassagegeo})
can be conveniently expressed in terms of the Schur measure, as
observed by K. Johansson in \cite{JohShapefluctuations}, see also
\cite{JohPNG}. He showed that \be \label{eqLnp} \mathbb{P}
(L(N,p)\leq n )=\frac{1}{Z}\sum_{\lambda_1 \leq n}
s_{\lambda}(a)s_{\lambda}(b). \ee This is a corollary of a more
general fact that the Schur measure with parameters $a$ and $b$ is
the image of the random integer valued matrix $W$ under the
Robinson-Schensted-Knuth correspondence, see
\cite{JohShapefluctuations}, \cite{JohPNG}.

\paragraph{}We now turn to the description
 of a continuous version of the Schur measure. Another treatment of
 the same object can be found in Appendix A to
 \cite{ForresterRains}.
Set $$a_i=1-\frac{\pi_i}{L},\ i=1, \ldots, p,\  \text{ and }\
b_j=1-\frac{\hat \pi_j}{L},\ j=1,\dots,N.$$ It is quite clear that
the distribution random variable $Y_L(N,p):=\frac{1}{L}L(N,p)$
should converge as $L \to \infty$ to that of the last passage time
in a percolation model with i.i.d. exponential random variables with
expectations $1/(\pi_i +\hat \pi_j)$. Define $Y(N,p)$ as in
(\ref{def: YNP}). Taking the limit $L \to \infty$ in (\ref{eqLnp})
and denoting $x_i=\lambda_i/L$, one readily obtains in the case where $N=p$ that
\begin{equation}
\mathbb{P}(Y(p,p)\leq x)= \frac{1}{Z_{p,p}}\int_{I^p}{\det \left
(e^{- \pi_i x_j} \right)}_{i,j=1}^p {\det \left (e^{- \hat \pi_i
x_j} \right)}_{i,j=1}^p\prod_{i=1}^p dx_i,\label{distriY}
\end{equation}
where $Z_{p,p}=\det \left (\frac{1}{\pi_i+\hat \pi_j}\right
)_{i,j=1}^p$ and $I=[0,x]$. The probability measure defined by the density \be
\label{def: densN}\frac{1}{Z_{p,p}}\det \left (e^{- \pi_i x_j}
\right)_{i,j=1}^p \det \left (e^{- \hat \pi_i x_j}
\right)_{i,j=1}^p\ee is the continuous version of the Schur measure.

The case $N<p$ can be handled via the limit transition $b_j\to 0$,
$j=N+1,\dots,p$, from the case $N=p$. The analog of (\ref{distriY})
reads
\begin{equation}
\mathbb{P}(Y(N,p)\leq x)= \frac{1}{Z_{N,p}}\int_{I^N}{\det \left
(f_{i}(x_j) \right)}_{i,j=1}^N {\det \left (e^{- \hat\pi_i
x_j}\right)}_{i,j=1}^N\prod_{i=1}^p dx_i,\label{distriY1}
\end{equation}
with
$$
f_{k}(x)=\frac 1{2\pi i}\oint
\frac{u^{k-1}e^{-ux}du}{\prod_{j=1}^p(u-\pi_j)}\,,\qquad
k=1,\dots,N,
$$
and the integration contour going around the poles
$\pi_1,\dots,\pi_p$.

The expression ${\det \left (f_i(x_j) \right)}_{i,j=1}^N$ can be
obtained as the limit of the Jacobi-Trudi formula for
$s_\lambda(a)$; it is also the limit, up to a constant, of the ratio
$$
\frac{\det {\left (e^{- \pi_i x_j}
\right)}_{i,j=1}^p}{\prod\limits_{N+1\le i<j\le p}(x_i-x_j)}
$$
as $x_{N+1},\dots,x_p$ converge to 0.

\section{A random matrix model associated to the continuous
version of the Schur measure.\label{Sec: RMT}}

The goal of this section is to prove Theorem \ref{theo: YNPRMT}.

\noindent Let $X_N$ be a $p \times N$ random matrix as in (\ref{def: XN}). Set
then $M_N=X_NX_N^*$.  Thus defined random matrix ensemble is a
natural generalization of the much studied complex Wishart ensemble.
In what follows we call it the {\it generalized Wishart ensemble\/}.
We show that the probability distribution of the largest eigenvalue
of $M_N$ has a density given by the right-hand side of (\ref{def:
densN}).

\paragraph{}Let us first consider the case where $p=N$.
 Similarly to the case of the ordinary Wishart ensemble, one
sees that the generalized Wishart ensemble is defined by the
probability density with respect to Lebesgue measure $L_p$ on the
space of complex matrices of size $p \times p$:
\be\label{density}\frac{d\mathbb{P}(X_p)}{dL_{p}}=const_p\,\exp{\{-\text{Tr}
S_1X_pX_p^*-\text{Tr} S_2X_p^*X_p\}}, \ee where
$S_1=\text{diag}(\pi_1, \ldots, \pi_p)$,  $S_2=\text{diag}(\hat
\pi_1, \ldots, \hat \pi_p),$ and $const_p$ is a positive constant.
Just as for complex Wishart ensembles with non identity covariance
matrix, the joint eigenvalue distribution of the generalized Wishart
ensembles can be explicitly computed.

Denote by $x_1\geq x_2\geq
\cdots \geq x_p$ the ordered eigenvalues of the sample covariance
matrix $X_pX_p^*$. Note that these eigenvalues are also the squared
singular values of $X_p$. Let $f(x_1, \ldots, x_p)$ denote the
density with respect to Lebesgue measure of the joint eigenvalue
distribution induced by the generalized Wishart ensemble. 
\bp
\label{Prop: jed}One has
$$
f(x_1, \ldots, x_p)= \frac 1{Z_{p,p}}\det \left ( e^{- \pi_i
x_j}\right)_{i,j=1}^p\det \left ( e^{-\hat\pi_{j}x_k}\right
)_{j,k=1}^p.
$$
\ep
\paragraph{Proof of Proposition \ref{Prop: jed}.}
Introduce the polar decomposition of the $p \times p$ matrix $X_p$:
One has
$$X=U D V\  \text{ with }\ U \in \mathbb{U}(p), \quad D=\text{diag}(\sqrt{x_1}, \ldots, \sqrt{x_p}), \text{ and } V
\in \mathbb{U}(p).$$ The joint eigenvalue distribution induced by
the probability measure (\ref{density}) can now be computed thanks
to the celebrated Itzykson-Zuber-Harisch-Chandra (IZHC) integral. We
have
$$f(x_1, \ldots, x_p)=const\cdot V(x)^2 \int_{\mathbb{U}(p)}e^{-\text{Tr} S_1 U D^2 U^*}dU \int_{\mathbb{U}(p)}e^{-\text{Tr} S_2 V D^2 V^*}dV.$$
In the above expression, $V(x)$ is the Vandermonde determinant:
$V(x)=\prod_{i<j}(x_i-x_j).$ The IZHC formula yields
$$\int_{\mathbb{U}(p)}e^{-\text{Tr} S_1 U D^2 U^*}dU
=\frac{\det \left ( e^{- \pi_i
x_j}\right)_{i,j=1}^p}{V(x)V(\pi)}\,.\qed$$

The proof that the largest eigenvalue of $X_NX_N^*$ has the same
distribution as the random variable $Y(N,p)$ can now be obtained
from Proposition \ref{Prop: jed} and formulas of the previous
section by the limit transition $\hat \pi_{N+1},
\dots,\hat\pi_p\to\infty$.

This finishes the proof of Theorem \ref{theo: YNPRMT}. \hfill $\square$

\section{A continuous version of the Schur Process \label{Sec: DRPF}}
In this section, we define a random point process which is a
continuous version of the Schur process introduced in \cite{Ok2}. We
consider the probability distribution on $\prod_{i=1}^p \mbR_+^i$
with density w.r.t. Lebesgue measure given by \be
\frac{1}{Z_{p}}\det \left ( e^{-\pi_i x_j^{p}}\right)_{i,j=1}^{p}
\prod_{k=1}^{p-1}\det \left ( e^{-*\hat \pi_{k+1}
(x_j^{k+1}-x_i^{k})}\right)_{i,j=1}^{k+1}e^{-\hat \pi_1
x_1^{1}}.\label{mult}\ee
Here we used the convention that
$x_{k+1}^{k}=0$ for any $1\leq k \leq p-1$, and the notation
$$e^{-*\hat \pi (x-y)}=\begin{cases} e^{-\hat \pi (x-y)},&x>y,\\
0,&\text{otherwise.}\end{cases}$$
The probability distribution (\ref{mult}) naturally arises here since the distribution of
$\max_{j=1,\dots,p}\{x_j^p\}$ is equal to the probability density function of $Y(p,p)$ (see Formula (\ref{distriY})).

\noindent Let $\mCC$ (resp. $\mCC'$) be a contour encircling the
$\{\pi_j\}_{j=1, \ldots, p}$ (resp. $\{-\hat \pi_j\}_{j=1, \ldots,
p}$) such that the two contours do not cross or contain each other.
Set
$$\Psi_{r,s}(u,v)=1_{r<s}1_{u<v}\:\frac{1}{2\pi i}\oint_{\mCC'}e^{w(v-u)}\prod_{k=r+1}^s\frac{1}{w+\hat \pi_k}dw.$$

The main result of this section is the following statement.

\bt \label{Prop: kernel} The random point process on
$\{1,\dots,p\}\times \mathbb{R}$ defined by the density (\ref{mult})
is determinantal, and its correlation kernel has the form

\begin{eqnarray}\label{kernel} 
&&
K(r, u;s,v)\crcr
&&=\frac{1}{(2\pi i)^2}\oint_{\mCC} dz
\oint_{\mCC'} dw \frac{e^{wv-zu}}{w-z} \dfrac{\prod_{k=1}^r (z+\hat
\pi_k)}{\prod_{l=1}^s(w+\hat \pi_l)}\prod_{i=1}^p
\frac{w-\pi_i}{z-\pi_i}-\Psi_{r,s}(u,v).\crcr
&& \end{eqnarray}
 \et

Let us briefly discuss the connection with the (discrete) Schur
process. A version of the Schur process has a natural interpretation
in terms of the last passage percolation model discussed in Section
\ref{Sec: RSK}. Let again $W=W^{(p)}$ be the $p\times p$ matrix
filled with geometrically distributed integers, and let $W^{(k)}$,
$k<p$, be the $p\times k$ matrix made of first $k$ columns of $W$.
Denote by $\lambda^{(k)}$ the image of $W^{(k)}$ under the
Robinson-Schensted-Knuth correspondence. Then the joint distribution
of $(\lambda^{(1)},\dots\lambda^{(p)})$ is given by the Schur
process: It has the form \be \label{densschur}const\cdot
s_{\lambda^{(p)}}(a_1, \ldots, a_p) s_{\lambda^{(p)}/
\lambda^{(p-1)}}(b_p)s_{\lambda^{(p-1)}/ \lambda^{(p-2)}}(b_{p-1})
\cdots s_{\lambda^{(2)}/ \lambda^{(1)}}(b_1),\ee where the notation
$s_{\lambda/\mu}$ stands for the skew Schur function. Considering
the case where the $a_i$'s and $b_j$'s approach $1$
($a_i=1-\pi_i/L$, $b_i=1-\hat \pi_j/L$ and $L \to \infty$), we can
then define the continuous limit of the Schur process, which leads
to (\ref{mult}). In the context of the percolation model, the
probability distribution (\ref{mult}) can be understood as the joint
distribution of the random Young diagrams obtained by the RSK
algorithm applied to matrices filled with independent but not
identically distributed exponential random variables; the
expectation of the $(i,j)$th entry is equal to $(\pi_i+\hat
\pi_j)^{-1}$.

Theorem \ref{Prop: kernel} could be derived from a limiting argument
for the correlation kernel of the Schur process, but we prefer to
give a self-contained random matrix oriented proof of Theorem
\ref{Prop: kernel} below.

The probability distribution (\ref{mult}) can also be viewed in the
random matrix theory context of the previous section. For $1\leq
k\leq p$, define $X_{k}$ to be the $p\times k$ matrix whose $k$
columns are the first $k$ columns of $X_p.$ Then, $M_k=X_kX_k^*$ is
a $p\times p$ random matrix of rank $k$. We denote by $x_i^{k},
1\leq i \leq k$ its nonzero eigenvalues. The formula (\ref{mult})
provides a good candidate for the joint distribution density of
$\{x_i^{k}\}$ in the sense that its projections to $\{x_i^{(k)}\}$
with fixed $k$ coincide with the densities of eigenvalues of $M_k$.
%Also, when all $\hat\pi_j$'s are equal, it is not hard to show that
%(the limiting value of) (\ref{mult}) is indeed the joint density of
%the eigenvalues of $M_k$.
Although we were unable to verify that (\ref{mult}) is indeed the
joint eigenvalue density for $(M_1,\dots,M_p)$, see the footnote on
page 3 for a reference to a partial result.

\paragraph{Proof of Theorem \ref{Prop: kernel}.}
We first consider an (algebraically simpler) auxiliary distribution
and then use an appropriate limit transition to compute the
correlation functions associated to (\ref{mult}).

\paragraph{}Instead of (\ref{mult}) let us
consider the probability distribution defined as follows. Let
$0<T_1<T_2<\cdots <T_{p-1}<T_p$ be positive numbers, and
$T_1=\hat\pi_1$. Define a probability distribution on
$(\mbR^p_+)^p=\{x_l^k\}_{k,l=1\dots,p}$ by the density
\be \label{multitime} \frac{1}{Z_p^T}\det \left (
\phi_{0,1}(x_0^i, x_1^j)\right)_{i,j=1}^p \prod_{r=1}^{p-1}\det
\left ( \phi_{r,r+1}(x_r^i, x_{r+1}^j)\right)_{i,j=1}^p\det \left (
\phi_{p,p+1}(x_p^i, x_{p+1}^j)\right)_{i,j=1}^p\ee where
$\phi_{0,1}(x_0^i, x)=e^{-*T_i x},$ $\phi_{r,r+1}(x,y)=e^{-*\hat
\pi_{r+1}(y-x)}$ and $\phi_{p,p+1}(x, x_{p+1}^j)=e^{-*\pi_j x}.$
Set
\begin{eqnarray*}
&\Psi_{0,s}^T(x_0^i, v)=&\int_{\mathbb{R}_+^{s-1}}
\phi_{0,1}(x_0^i,x_1)\crcr&&
\left(\prod_{k=1}^{s-2}\phi_{k,k+1}(x_k,
x_{k+1})\right)\! \phi_{s-1,s}(x_{s-1},v)\prod_{i=1}^{s-1}dx_i,\crcr
&\Psi_{r,p+1}(u, x_{p+1}^j)=&\int_{\mathbb{R}_+^{p-r}}
\phi_{r,r+1}(u,x_{r+1})\crcr
&&\left(\prod_{k=r+1}^{p-1}\phi_{k,k+1}(x_k,
x_{k+1})\right)\phi_{p,p+1}(x_{p},x_{p+1}^j)\prod_{i=r+1}^{p}dx_i,\crcr 
&\Psi_{r,s}(u,v)=&1_{r<s}\int_{\mathbb{R}_+^{s-r}}
\phi_{r,r+1}(u,x_r)\crcr
&&\left(\prod_{k=r+1}^{s-2}\phi_{k,k+1}(x_k,x_{k+1})
\right)\phi_{s-1,s}(x_{s-1},v)\prod_{i=r}^{s-1}dx_i.
\end{eqnarray*}

\bl\label{l:explicit} For any $1\le i,j\le p$ and $u,v>0$ we have
\begin{eqnarray*}  &&\Psi_{0,s}^T(x_0^i, v)=\frac{1}{2\pi i}\oint
\frac{e^{wv}}{w+T_i}\prod_{k=2}^s\frac{1}{w+\hat \pi_k}\,dw,\crcr
&&\Psi_{r,p+1}(u, x_{p+1}^j)=e^{-\pi_j u}\prod_{k=r+1}^p
\frac{1}{\pi_j+\hat \pi_k}\,,\crcr
&&A_{ij}:=\int_0^{+\infty} \Psi_{0,s}^T(x_0^i, u)\Psi_{r,p+1}(u,
\pi_j)du=\frac{1}{T_i+\pi_j}\prod_{k=2}^p \frac{1}{\pi_j+\hat
\pi_k}\,,\crcr
&&\Psi_{r,s}(u,v)=1_{r<s}1_{u<v}\:\frac{1}{2\pi i}\oint
e^{w(v-u)}\prod_{k=r+1}^s\frac{1}{w+\hat \pi_k}dw.
\end{eqnarray*}
The integration contours are positively oriented loops that contain
all poles of the integrands.\el 
The proof of Lemma \ref{l:explicit} consists of inductions on $s$ for $\Psi_{0,s}^T(x_0^i, v)$, on
$(p-r)$ for $\Psi_{r,p+1}(u, x_{p+1}^j)$, and on $(s-r)$ for
$\Psi_{r,s}(u,v)$. The formula for $A_{ij}$ is proved by a
straightforward residue computation.

Let us now apply the Eynard-Mehta theorem (see \cite{EynardMehta},
\cite{NagaoForrester}, \cite{TW-Dyson}, \cite{JohPNG},
\cite{BorodinRains}) to compute the correlation functions of
(\ref{multitime}). For $1\le r,s\le p$, denote
\begin{eqnarray*}&\Psi_{r,s}^T(u,v)=&1_{r<s}\int_{\mathbb{R}_+^{s-r}}
\phi_{r,r+1}(u,x_r)\crcr
 &&\left(\prod_{k=r+1}^{s-2}\phi_{k,k+1}(x_k,x_{k+1})
\right)\phi_{s-1,s}(x_{s-1},v)\prod_{i=r}^{s-1}dx_i.
\end{eqnarray*}

\bp \label{Prop: corrkmultitime}The random point process on
$\{1,\dots,p\}\times \mathbb{R}_+$ defined by the measure
(\ref{multitime}) is determinantal, and its correlation kernel can
be written in the form 
\begin{eqnarray} &K_T(r, u;s,v)=&\frac{1}{(2\pi
i)^2}\oint_{\mCC_1} dz \oint_{\mCC'_1} dw \frac{e^{wv-zu}}{w-z}\crcr
&&\dfrac{\prod_{k=2}^r (z+\hat \pi_k)}{\prod_{k=2}^s(w+\hat
\pi_k)}\prod_{i=1}^p
\frac{(w-\pi_i)(z+T_i)}{(z-\pi_i)(w+T_i)}-\Psi_{r,s}(u,v),\label{intrepT}
\end{eqnarray} 
where the contour $\mCC_1$ encircles the $\pi_j, j=1, \ldots,
p$, the contour $\mCC_1'$ encircles the $-\hat \pi_j, -T_j,$ for $
j=1, \ldots, p$, and the two contours do not cross or contain each
other. \ep
\paragraph{Proof of Proposition \ref {Prop: corrkmultitime}:}
The Eynard-Mehta theorem implies that the random point process in
question is determinantal, and that its correlation kernel can be
written as
\begin{equation*}
K_T(r,u;s,v)=\sum_{i,j=1}^p
\Psi_{r,p+1}(u, x_{p+1}^i) A^{-1}_{ij} \Psi^T_{0,s}(x_0^j,
v)-\Psi_{r,s}(u,v).
\end{equation*}
 Using the formula for the
determinant of the Cauchy matrix one explicitly computes
$A_{ij}^{-1}$, which together with the formula for $\Psi_{0,s}^T$
from Lemma \ref{l:explicit} yields
\begin{eqnarray*}&&K_T(r,u;s,v)+ \Psi_{r,s}(u,v)=\crcr
 &&\sum_{i=1}^p
\Psi_{r,p+1}(u, x_{p+1}^i)\frac{1}{2\pi i}\oint e^{wv}
\frac{\prod_{k=2}^p (\pi_i+\hat \pi_k)}{\prod_{k=2}^s(w+\hat
\pi_k)}\prod_{j=1}^p \frac{\pi_i+T_j}{w+T_j} \prod_{k\not=
i}\frac{w-\pi_k}{\pi_i-\pi_k}dw,
\end{eqnarray*}
where the contour contains all poles of the integrand. A final
residue computation  yields the integral expression
(\ref{intrepT}).\hfill $\square$

\paragraph{}We can now come back to the computation of correlation functions for the probability distribution (\ref{mult}) and finish the proof of Theorem \ref{Prop: kernel}.
The probability distribution (\ref{mult}) can be obtained from
(\ref{multitime}) by taking the limit $T_{p}>T_{p-1}>\cdots >T_2 \to
\infty.$ The proof of Theorem \ref{Prop: kernel} is now a
straightforward corollary of Proposition \ref{Prop: corrkmultitime}.
 \hfill $\square$
\section{\label{Sec: asym}An extension of the Airy point process}
In this section, we first consider the case where $\hat \pi_i=0$ and
$\pi_i=1$ for any $i=1, \ldots, p$. We then show that in this case,
the suitably rescaled correlation functions converge in the
large-$p$-limit to those defined by the extended Airy kernel. Then,
to define a new extended Airy-type kernel with parameters, we will
allow a certain number of these parameters to depend on $p$ and
study the rescaled correlation functions.
\subsection{\label{subsec: extenAiry}The simple case $\hat \pi_i=0$ and $\pi_i=1$ for any $i$: the extended Airy kernel.}
The extended Airy kernel is an extension of the well-known Airy
kernel; it occurs for example as the limiting correlation kernel for
the process of largest eigenvalues of Dyson's Brownian Motion on
Hermitian matrices.

Lemmas 2 and 3 proved in this section are also a part of Proposition
5 of \cite{ForresterNagao}.

%\bdefi
%The extended Airy kernel is defined by the integral representation:
%\begin{eqnarray*}
%&&K_{Ai}(t_1,x;t_2,y):=\int_{0}^{\infty}
%e^{-\lambda(t_1-t_2)}Ai(y+\lambda)Ai(x+\lambda)d\lambda\ \text{, if
%}\ t_1\ge t_2,\crcr &&K_{Ai}(t_1,x;t_2,y):=-\int_{-\infty}^{0}
%e^{-\lambda(t_1-t_2)}Ai(y+\lambda)Ai(x+\lambda)d\lambda\ \text{, if
%}\ t_1 <t_2.
%\end{eqnarray*}
%\edefi
Let $K_{Ai}(t_1,x;t_2,y)$ be the extended Airy kernel defined in (\ref{def: airykernelextended}).
Here we show that, when suitably rescaled, the asymptotics of the correlation kernel
\begin{eqnarray}
 \label{tildekp} &K_p(r, u;s,v)=&\frac{p}{(2\pi i)^2}
 \oint_{\mCC_1} dz \oint_{\mCC'_1} dw\frac{e^{pwv-pzu}}{w-z} \crcr
&&
\dfrac{\prod_{k=1}^r (z+\hat \pi_k)}{\prod_{k=1}^s(w+\hat
\pi_k)}\prod_{i=1}^p \frac{w-\pi_i}{z-\pi_i}-p\Psi_{r,s}(pu,pv)\,,
\end{eqnarray}
which is a rescaled version of (\ref{kernel}), is given by the extended Airy kernel.\\
Due to the choice of the $\pi_i$'s, and  $\hat \pi_j$'s, one can
write that
\begin{eqnarray*}
&\tilde K_p(r, u;s,v):&= K_p(r,
u;s,v)+p\Psi_{r,s}(pu,pv)\crcr
&&=\frac{p}{(2i\pi)^2}\int_{\mCC_1} dz
\int_{\mCC'_1} dw \,e^{p\,(F_{v,s}(w)-F_{u,r}(z))}\frac{1}{w-z}\,,
\end{eqnarray*}
where $F_{u,r}(z)=uz+\ln (z-1)- \frac{r}{p}\ln z.$

Let $0<t<1$ be some given real number independent of $p.$ Let also $t_1, t_2$ be given.
In the following, we set
\begin{gather}
r=\left[tp+p^{2/3}\dfrac{2\sqrt{t}(1+\sqrt{t})^2}{\alpha}t_1\right]:=s_1p,\label{def: rs}
\\ s=\left[tp+p^{2/3}\dfrac{2\sqrt{t}(1+\sqrt{t})^2}{\alpha}t_2\right]:=s_2p,\\
%\alpha=\dfrac{(1+\sqrt{t})^{4/3}}{t^{1/6}},
%\label{def: alpha}\\
\alpha=\dfrac{(1+\sqrt{t})^{4/3}}{t^{1/6}},\quad u=(1+\sqrt{s_1})^2
+\dfrac{\alpha x}{p^{2/3}},\quad v=(1+\sqrt{s_2})^2 +\dfrac{\alpha
y}{p^{2/3}}.\label{rescalings}
\end{gather}
%Note that $\dfrac{\alpha_{s_i} }{p^{2/3}}=\dfrac{\alpha
%}{p^{2/3}}+O(p^{-1})$.
Here $[x]$ stands for the integral part of $x\in \mathbb{R}$.

We first consider the case where $s\leq r.$ Set $\Gamma:=\{ te^{\pm
2i\pi/3}, t\in \mathbb{R}_+\}$ to be a contour oriented from bottom
to top and $\gamma:=\{te^{\pm i\pi/3},t\in \mathbb{R}_+\}$ to be
oriented from top to bottom.

\bl\label{Lem: Airy} With the above rescaling, for $s\leq r$ and
$z_0:=\frac{\sqrt{t}}{1+\sqrt{t}}$ we have
\begin{eqnarray*}
&&\displaystyle{\lim_{p \to
\infty}\frac{\alpha}{p^{2/3}}}\,e^{p\,(F_{u,r}(z_0)-F_{v,s}(z_0))}\tilde
K_p(r, u;s,v)\crcr
&&=\frac{e^{yt_2-xt_1+\frac{1}{3}(t_1^3-t_2^3)}}{(2\pi
i)^2}\int_{\gamma}ds'\int_{\Gamma}dt'
\,\frac{e^{yt'-\frac{t'^3}{3}-xs'+\frac{s'^3}{3}}}{t'-s'+t_2-t_1}\crcr
&&=e^{yt_2-xt_1+\frac{1}{3}(t_1^3-t_2^3)}\int_{0}^{\infty}
e^{-\lambda(t_1-t_2)}Ai(y+\lambda)Ai(x+\lambda)d\lambda.\crcr &&
\end{eqnarray*}
\el
\paragraph{Proof of Lemma \ref{Lem: Airy}.}
It is convenient to define $u_o=(1+\sqrt{s_1})^2 \text{ and
}v_o=(1+\sqrt{s_2})^2.$ The reason for the above rescaling
(\ref{def: rs})--(\ref{rescalings}) is that
\begin{eqnarray*}&F_{u,r}(z)&=F_{u_o,r}(z)(1+o(1))\crcr&&= (u_oz+\ln (z-1)-s_1\ln z)(1+o(1)):=f_{s_1}(z)(1+o(1)),\end{eqnarray*}
where the function $f_{s_1}(z)$ has a degenerate critical point at $$z_c:=\frac{\sqrt{s_1}}{1+\sqrt s_1}\sim z_0+O(p^{-1/3}).$$
In particular, one has that:
$$f_{s_1}'(z_c)=f_{s_1}''(z_c)=0, \text{ and } f_{s_1}'''(z_c)=\frac{-2(1+\sqrt{s_1})^4}{\sqrt{s_1}}.$$
To obtain the leading term in the asymptotic expansion of $K_p$, we
define the following contours. Using the notation
$w_c=\frac{\sqrt{s_2}}{1+\sqrt{s_2}}$, set
$$\mCC_{1,1}=\{z_c+s'e^{\pm i\pi/3},\: 0\leq s'<\delta_o\},\:\mCC'_{1,1}=\{w_c+t'e^{\pm i2\pi/3},\: 0\leq
t'<\delta_1\},\label{contours1}$$
where $\delta_o>0$ and $\delta_1>0$ are constants that will be determined in the sequel.
These contours are completed as follows. Set $\theta_o=\text{arg}(z_c+\delta_oe^{i\pi/3})$ (resp. $\theta_1=\text{arg}(w_c+\delta_1e^{2i\pi/3})$), where $\text{arg}$ denotes the argument of a complex number. Set
\begin{eqnarray*}&&\mCC'_{1,2}=\{|w_c+\delta_1e^{2i\pi/3}|e^{i \theta}, \theta_1\leq \theta \leq 2\pi-\theta_1]\},\crcr &&\mCC_{1,2}=\{1+|z_c+\delta_oe^{i\pi/3}-1|e^{i \theta},  |\theta|\leq \theta_o\}.\end{eqnarray*}
The constant $\delta_o$ is also large enough so that the $z$-contour
encircles all the $\pi_i$'s (even in the case where some of them
differ from $1$).
\vspace*{-0.1cm}\begin{figure}[htbp]
 \begin{center}
 \begin{tabular}{c}
 \epsfig{figure=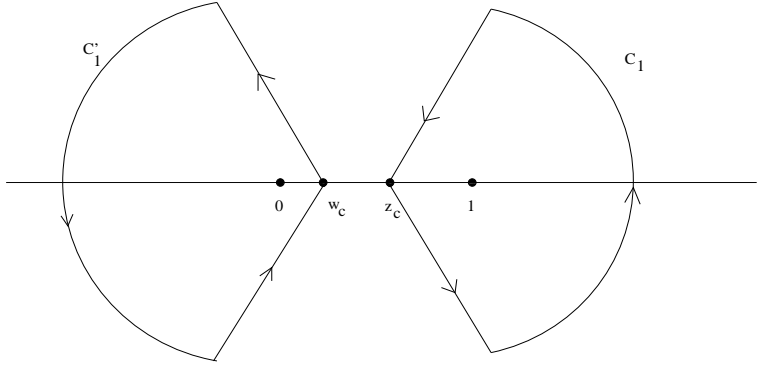,height=4.8cm,width=10cm, angle=0}
 \end{tabular}
 \caption{The two contours $\mathcal{C}_1$ and $\mathcal{C}'_1$ \label{fig: contourcasdebile}}
\vspace*{-0.3cm}
 \end{center}
\end{figure}
It can then be checked that
$$\frac{\partial}{\partial s}\Re\left(f_{s_2}(w_c+\frac{se^{2i\pi/3}}{1+\sqrt s_2})\right)=-\frac{s^2(1+\sqrt{s_2})(s^2-s(1-\sqrt{s_2})+2\sqrt{s_2})}{2(s^2+s+1)(s^2-s\sqrt{s_2}+s_2)}<0 
$$
for any $s>0$ provided $\sqrt{s_2}>5-\sqrt{96}/2$. If $\sqrt{s_2}<5-\sqrt{96}/2$, we set $\delta_1$ to be the smallest positive root of the polynomial $X^2-X(1-\sqrt{s_2})+2\sqrt{s_2}.$ Otherwise $\delta_1$ is arbitrarily large.
Similarly
$$\frac{\partial}{\partial s}\Re
\left(f_{s_1}(z_c+\frac{se^{i\pi/3}}{1+\sqrt
s_1})\right)=\frac{s^2(1+\sqrt{s_1})(s^2+s(1-\sqrt{s_1})+2\sqrt{s_1})}{2(s^2-s+1)(s^2+s\sqrt{s_1}+s_1)}>0,$$
for any $s>0$. We also have that if $|\delta|$ is bounded
$$\left|pf_{s_2}\left(w_c+\frac{\delta}{p^{1/3}}\right)-pf_{s_2}(w_c)-f_{s_2}^{(3)}(w_c)
\frac{\delta^{3}}{3!}\right|\leq
\frac{\sup_{\Omega}|f_{s_1}^{(4)}(w)|}{p^{1/3}}\leq C\frac{\delta^4}{p^{1/3}}.$$
Here $w=w_c+\frac{\delta}{p^{1/3}}$ lies in a compact subset $\Omega$ of $\mbC\setminus\{0,1\}.$\\
To complete the proof one needs to verify that on remaining parts of
the contours the integrand becomes exponentially small as $p$ gets
large. If we set
$w:=|w_c+\delta_1e^{2i\pi/3}|e^{i \theta}$, and using the fact that $v_o=|w_c-1|^{-2}$, one can check that
$$\Re\left(\frac{\partial }{\partial \theta}f_{s_2}(w)\right)= -\Im (w) \left (v_o-\frac{1}{|w-1|^2}\right )<0,$$
if $\theta \in [\theta_1, \pi].$ It is an easy
computation to check the remaining parts of the contours, and we
omit it.

If we assume that $t_1, t_2, x, y$ lie in a compact set of $\mbR$,
then the above estimates imply that
\begin{itemize}
\item It is enough to integrate over a neighborhood of radius $p^{-1/3}p^{1/12-\epsilon}$ of the critical points.
\item Inside such a neighborhood the Taylor expansion holds.
\end{itemize}

 Writing out
this expansion explicitly yields Lemma \ref{Lem: Airy}. \hfill
$\square$

In the case $s>r$ we cannot make the integration contours go through
their corresponding critical points so that they do not intersect.
We then modify the contours in a neighborhood of width $p^{-1/3}$ of
$z_c$ and $w_c$ so that the $w-$contour remains on the left of the
$z-$contour. This does not modify the saddle point argument. We just
need to consider the function $\Psi_{r,s}$ separately, which is done
below.

\paragraph{}
Due to the rescaling of the correlation kernel, in the case where $r<s$, we need to consider
the asymptotics of
$$\frac{\alpha}{p^{2/3}}e^{p(F_{u,r}(z_0)-F_{v,s}(z_0))}
\frac{p}{2\pi i}\oint_{\gamma_o} e^{pw(v-u)}\frac{dw}{w^{s-r}},$$
where $\gamma_o$ is a contour encircling the pole $w=0.$

\bl \label{Lem: psirs} For $s>r$, using the scaling (\ref{def:
rs})--(\ref{rescalings}) one has
\begin{eqnarray*}&&\lim_{p \to \infty}\frac{\alpha}{p^{2/3}}e^{pF_{u,r}(z_0)-pF_{v,s}(z_0)}
K_p(r,u;s,v)\crcr&&=-e^{yt_2-xt_1+\frac{(t_1^3-t_2^3)}{3}}\int^{0}_{-\infty}
e^{-\lambda(t_1-t_2)}Ai(y+\lambda)Ai(x+\lambda)d\lambda.\end{eqnarray*} \el
\paragraph{Proof of Lemma \ref{Lem: psirs}:}Setting
$\beta= {2\sqrt t (1+\sqrt t)^2}/{\alpha},$ one has
\begin{eqnarray*}
&&\frac{\alpha}{p^{2/3}}e^{\{pF_{u,r}(z_0)-pF_{v,s}(z_0)\}}\frac{p}{2\pi
i}\oint_{\gamma_o} e^{pw(v-u)}\frac{dw}{w^{s-r}}\crcr
&&=\frac{\alpha}{p^{2/3}}e^{\{pF_{u,r}(z_0)-pF_{v,s}(z_0)\}}\frac{p}{2
\pi i}\oint_{\gamma_o} dw\frac{\exp{\left\{p^{2/3}\frac{(1+\sqrt{t})\beta
(t_2-t_1)}{\sqrt t}w\right\}}}{w^{\beta p^{2/3}(t_2-t_1)+O(1)}}
\crcr &&\times \exp{\left\{p^{1/3}\left (-\frac{\beta^2}{4t^{3/2}}
(t_2^2-t_1^2)
w+\alpha(y-x)w+o(1)\right)\right\}}.\end{eqnarray*}

Consider
$$F(w)=\beta (t_2-t_1)\frac{1+\sqrt{t}}{\sqrt t}w-\beta (t_2-t_1)\ln
w.$$ It is not hard to see that the critical point of this function
is $w_0=z_0=\frac{\sqrt{t}}{1+\sqrt{t}}$ and $F''(z_0)=\beta
(t_2-t_1){z_0^{-2}}.$ A contour which satisfies the saddle point
analysis requirement can be chosen as follows: $\gamma_o=\gamma_1
\cup \overline{\gamma_1}$ where $\gamma_1=\{z_0+it, |t|\leq
z_0\}\cup \{z_0e^{i(\pi/4+\theta)}, 0<\theta<3\pi/4\}$.

We obtain
\begin{eqnarray*}
&&\quad\lim_{p \to
\infty}\frac{\alpha}{p^{2/3}}\exp{\{pF_{u,r}(z_0)-pF_{v,s}(z_0)\}}\frac{p}{2\pi
i}\oint_{\gamma_o} e^{pw(v-u)}\frac{dw}{w^{s-r}}\crcr 
&&=\lim_{p \to\infty}\frac{\alpha }{2 \pi i } \frac{1}{\sqrt{F''(z_0)}}\int_{i
\mbR}\exp{\left\{\frac{s'\left(
\alpha(y-x)-\frac{\beta^2(t_2^2-t_1^2)}{4t^{3/2}}\right)}{\sqrt{F''(z_0)}}+\frac{s'^2}{2}\right\}}ds'\crcr
&&=\frac{1}{2\pi i\sqrt{2(t_2-t_1)}}\int_{i\mbR}\exp{\left
\{\frac{s'^2}{2}+\frac{(y-x)s'-(t_2^2-t_1^2)s'}{\sqrt{2(t_2-t_1)}}\right\}}ds'\crcr\label{asym:
psirs}&&=\frac {1}{\sqrt{4\pi (t_2-t_1)}}\,\exp\left\{
-\frac{(y-x+t_1^2-t_2^2)^2}{t_2-t_1}\right\}.
\end{eqnarray*}

Proposition 2.3 of \cite{JohPNG} completes the proof. \hfill
$\square$
\subsection{Extended Airy kernel with two sets of parameters}
We now consider the case where some of the $\pi_i$'s (resp. $\hat
\pi_j$'s) differ from $1$ (resp. $0$). This allows us to obtain a
new extended Airy type kernel with
two sets of parameters and prove Theorem 2.\\
We assume that (\ref{def: param}) holds true and 
%we consider the case 
%where there exist two integers $J_1, J_2$
%independent of $p$ such that:
%\begin{eqnarray*}
%&\pi_i=\dfrac{\sqrt{t}}{\sqrt t +1 }+\dfrac{x_i}{\alpha
%p^{1/3}},\quad i=1, \ldots, J_1;\quad& \hat
%\pi_j=-\dfrac{\sqrt{t}}{\sqrt t +1 }-\dfrac{y_j}{\alpha
%p^{1/3}},\quad  j=1, \ldots, J_2.\label{def: param}
%\end{eqnarray*}
%Here we will assume 
that all the $x_i$'s and $y_i$'s lie in a fixed
compact set of $\mathbb{R}$. We also assume that $ x_i-y_j
>0$ for any $i,j$, so that the joint distribution (\ref{mult})
 is well defined.
%We also assume that the distinguished parameters $\pi_i  \not=1$ and $\pi_j \not=0$ are in the ``first'' columns and rows of the matrix. This is a technical restriction which
%simplifies the statement of the results, but does not impact on the
%limiting kernel.
%This condition ensures essentially that, for $r$
%and $s$ large enough, \be \mathcal{J}(r):=\{ i\leq r,\hat \pi_i
%\not=0\}=\mathcal{J}(s)=\{\hat \pi_i
%\not=0\}:=\mathcal{J}.\label{defdeJ} \ee
%Define then
%$$\mathcal{I}:=\{  i\leq p, \pi_i \not=1\}\text{ and } I:=\sharp
%\mathcal{I}.$$

 \bt \label{Prop: extendedAiryparam1} With the above
rescaling, one has
\begin{eqnarray*}
&&\lim_{p \to
\infty}\frac{\alpha}{p^{2/3}}e^{pF_{u,r}(z_0)-pF_{v,s}(z_0)} K_p(r,
u;s,v)=(\ref{extparaAiry2})
\end{eqnarray*}
where the integration contours $\gamma$ and $\Gamma$ are chosen as
in Figure \ref{fig: contoursdefi}. \et
\brem Theorem \ref{Prop: extendedAiryparam1} readily implies Theorem \ref{theo: CVext}.
\erem 
\paragraph{Proof of Theorem \ref{Prop: extendedAiryparam1}:} The proof
relies on the same saddle point analysis of the correlation kernel
as in the previous section. In the expression (\ref{tildekp}) we
replace
$$
\frac 1{w-z}\dfrac{\prod_{k=1}^r (z+\hat \pi_k)}{\prod_{k=1}^s(w+\hat
\pi_k)}\prod_{i=1}^p \frac{w-\pi_i}{z-\pi_i}
$$
by
$$
\dfrac{\prod_{k=J_2+1}^r (z+\hat \pi_k)}{\prod_{k=J_2+1}^s(w+\hat
\pi_k)}\prod_{i=J_1+1}^p \frac{w-\pi_i}{z-\pi_i}\left(\frac
1{w-z}+\frac {\left(\prod_{k=1}^{J_2} \frac{z+\hat
\pi_k}{w+\hat \pi_k}\prod_{i=1}^{J_1}
\frac{w-\pi_i}{z-\pi_i}-1\right)}{w-z}\right)
$$
and observe that the second summand has no singularity at $z=w$.
This allows us to use the same contour deformation as in the
previous section, which directly leads to the result. \qed

\end{document}